\documentclass[letterpaper]{article} 
\usepackage[preprint]{aaai2027}  
\usepackage[hyphens]{url}  
\usepackage{graphicx} 
\usepackage{natbib}  
\usepackage{caption} 
\usepackage{algorithm}
\usepackage{algorithmic}

\usepackage{newfloat}
\usepackage{listings}
\DeclareCaptionStyle{ruled}{labelfont=normalfont,labelsep=colon,strut=off} 
\floatstyle{ruled}
\newfloat{listing}{tb}{lst}{}
\floatname{listing}{Listing}

\usepackage{booktabs}
\usepackage{array}
\usepackage{amsmath}

\usepackage{xurl}

\title{Living off the Local Model: Abusing Browser-Embedded\\Language Models for Post-Exploitation}

\author{
    William Fleshman\textsuperscript{\rm 1},
    Joseph Lucas\textsuperscript{\rm 2}
}
\affiliations{
    \textsuperscript{\rm 1}Army Cyber Institute\\
    \textsuperscript{\rm 2}NVIDIA\\
    william.fleshman@westpoint.edu, jolucas@nvidia.com
}

\begin{document}

\maketitle

\begin{abstract}
Browsers now ship local language models to ordinary endpoints. Google includes Gemini Nano in Chrome to enable productivity features such as summarization, with inference running entirely on the device. These browser-embedded models provide new \emph{living-off-the-land} opportunities for attackers, each instance a potential post-exploitation engine already installed, signed by the vendor, and invisible to network monitoring. The Gemini model is multimodal, capable of processing files, analyzing screenshots, or conversing with an attacker. We show that it can be driven from the host using only interpreters that ship with the operating system and find that the one-time user-activation check before the model is downloaded is easily satisfied with a simulated click or bypassed entirely. We measure and report the current model's ability to identify credentials on the host, detect PII in email, classify images, and call custom tools. We combine these capabilities and use the model to carry out a simulated cyber operation: reading a password from a webcam image, unlocking an encrypted archive, triaging the contents, and staging a private key for exfiltration, end-to-end without user-interaction. 

\end{abstract}

\section{Introduction}

Living off the land (LotL) is a well-known post-exploitation tradecraft where an intruder works with the trusted tools already on the accessed device~\cite{cisa_lotl}. These might include PowerShell, Python, WMI, or the long list of binaries, scripts, and libraries documented by the LOLBAS project~\cite{lolbas}. The proliferation of large language models (LLMs) makes such intrusions more adaptive, but the size of these models has generally limited LLM-enabled malware to depend on remote APIs, requiring credentials the provider can revoke, and the generation of network traffic to well-known endpoints that defenders can monitor~\cite{oesch2026living}. 

Small language models (LMs) remove these dependencies. Chrome now downloads a 2--4 GB multimodal LM to desktops, Gemini Nano, and exposes it to web content through the Prompt API~\cite{chrome_builtin_ai_getstarted, chrome_prompt_api}. Microsoft Edge is adding equivalent built-in APIs backed by an on-device Phi model, and Windows exposes a local Phi model to apps on Copilot+ PCs~\cite{windows_phi_silica,edge_prompt_api}. For an attacker, these already present offline models are exactly the type of tools worth leveraging toward malicious ends.

\begin{figure}[t]
  \centering
  \includegraphics[width=\columnwidth]{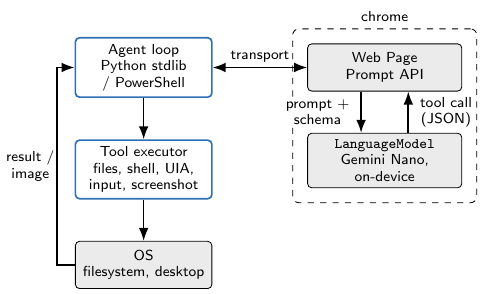}
  \caption{Agent architecture. Shaded components ship with the OS or with Chrome;
  the attacker supplies only an interpreter script. The agent loop prompts the
  on-device model through a web page that calls the Prompt API.}
  \label{fig:arch}
\end{figure}

\citet{oesch2026living} recently introduced \emph{Living off the LLM}, a paradigm for attackers to opportunistically use a victim's self-deployed model infrastructure~\cite{oesch2026living}. We study the more ubiquitous case: small models broadly deployed by trusted vendors such as Google or Microsoft. Specifically, we experiment with using Chrome's embedded Gemini Nano model as a post-exploitation agent (Figure~\ref{fig:arch}).

In this paper, we:
\begin{itemize}
    \item Describe the threat model for LotL use of browser-embedded models;
    \item Show how to satisfy or bypass model gating as an attacker;
    \item Measure the capability of Chrome's Gemini Nano model to perform common post-exploitation tasks;
    \item Build agentic harnesses around the model using OS-native interpreters such as PowerShell;
    \item Demonstrate the model's ability to autonomously conduct a multi-step cyber operation; and
    \item Provide mitigations and recommendations for defenders.
\end{itemize}

\section{Background and Related Work}

\subsection{Living off the Land (LotL)} Attackers use LotL techniques to abuse legitimate, pre-installed software such as interpreters, administration tools, and system binaries so that malicious activity blends into normal operations and to prevent dropped artifacts from triggering security products~\cite{cisa_lotl}. For example, \citet{barrsmith2021survivalism} systematically surveys LotL use in a broad set of Windows malware~\cite{barrsmith2021survivalism}. The MITRE ATT\&CK framework covers multiple techniques falling under this pattern such as Command and Scripting Interpreter (T1059) and System Binary Proxy Execution (T1218)~\cite{mitre_attack}. Because these tools are inherently dual-use, defenders must rely on behavioral indicators rather than signature-based approaches.

\subsection{Local and LLM-Enabled Offense} LLMs have already been evaluated as penetration-testing assistants~\cite{deng2024pentestgpt}, and traditional offensive techniques like \emph{worms} have been adapted against assistant ecosystems~\cite{cohen2024worm}. Closest to our work, \citet{oesch2026living} argue that local models will become LotL assets, and they demonstrate the idea using an Ollama-hosted model~\cite{oesch2026living}. In this work, we explore vendor-deployed browser models that users may not even be aware exist on their device.

\subsection{Gemini Nano} We specifically study Chrome's built-in Gemini Nano model, which exposes pages through the \texttt{LanguageModel} JavaScript interface~\cite{chrome_prompt_api}. A page first checks \texttt{availability()} (one of \emph{unavailable}, \emph{downloadable}, \emph{downloading}, or \emph{available}), then creates a session, and prompts the model. Sessions can accept text or image input, and prompts can constrain the model output to a passed JSON schema, which we leverage to enable tool-calling. The API became available with Chrome 148 on Windows 10/11, macOS 13+, Linux, and ChromeOS~\cite{chrome_builtin_ai_getstarted}. If \emph{available}, the model is downloaded the first time an origin uses the API, requiring a user gesture to start the download. Then the model will be \emph{available} for use without additional gestures. Enterprise policies can disable the APIs or the model as described in the Mitigations section.

\subsection{Browser Agents and Attacks} Previous attacks on LMs built into the browser come from extensions and web pages. CVE-2026-0628 let an extension with basic permissions hijack Chrome's Gemini Live panel and reach the camera, microphone, screenshots, and local files~\cite{unit42_gemini_live}. An RSAC 2026 talk showed how web pages and extensions can abuse the same Gemini model and API we leverage from the host in this work~\cite{rsac2026_prompt_to_pwn}. On the privacy side, whether on-device APIs are available, or how fast they respond, can reveal a device's hardware class to an attacker~\cite{datadome2026fingerprint}, similar to the timing leaks measured in hosted LLM APIs~\cite{gu2025auditing}. Benchmarks such as OSWorld evaluate agents that operate desktops through screenshots and accessibility data~\cite{xie2024osworld}. Our agent follows a similar design with a far smaller model.

\section{Threat Model}
We consider an attacker who has gained access to a machine as an ordinary user, for example, through phishing, and who now wants to perform adaptive post-exploitation (reconnaissance, semantic search, staging, etc.) without generating traffic to LLM providers. The attacker may use only what is already present: OS-native interpreters, the Chrome binary, and the on-device model (or the ability to download the model using our disclosed techniques). 

This attacker is excluded from Chrome's existing threat model: the Chrome Security FAQ notes that code running as the user ``can do anything you can do''~\cite{chromium_security_faq} so their vulnerability rewards program excludes bugs relying on existing access~\cite{chrome_vrp_faq}. This incentivizes attackers to use Chrome as a LotL asset, as reported issues might not get fixed. The consequences of our threat model therefore fall to the organization's set policies or endpoint monitoring, which is where our recommendations are aimed.

\section{Techniques}
\label{sec:technique}

\subsection{Gaining Model Access}

The Chrome DevTools Protocol (CDP) is the interface that developer tools and popular automation frameworks such as Puppeteer and Playwright use to control the browser~\cite{cdp}. CDP is started with a remote-debugging flag, Chrome then lets a local client evaluate JavaScript in the active page and dispatch events the page treats as trusted. Beginning with Chrome 136, the flag is disabled for the default profile, a change Google made after infostealers used the CDP to extract cookies~\cite{chrome_remote_debugging_136}. Now it requires a separate profile directory, which is no obstacle under our threat model and assumed access.

Our clients only use software already on the host, such as PowerShell or Python with standard libraries only. A script launches Chrome (can be headless) with a remote-debugging port and a dedicated profile directory, opens a page, and connects to that page's DevTools endpoint over a loopback WebSocket. Our Linux client implements the WebSocket in Python using the \texttt{socket} module while our Windows client uses PowerShell with .NET's builtin \texttt{ClientWebSocket}. Through CDP's \texttt{Runtime.evaluate}, the client calls \texttt{availability()}, creates a session, prompts the model, and receives the model reply as the return value. If the model is not yet present in the profile, the client first sends a synthetic mouse press and release with \texttt{Input.dispatchMouseEvent}, which we found satisfies the required user gesture to trigger the initial model download. Alternatively, we found that if an existing profile has already downloaded the model, we can bypass the user gesture by simply copying the model files and \texttt{Local State} from that profile to our own. We disclosed both user-activation gate bypasses to Google~\cite{chromium_issue_583903270}; the report was marked \emph{Won't Fix}, consistent with the threat-model exclusion discussed in the previous section.

The debugging protocol is a convenience, but not a requirement for the attacker. The same model is reachable from an ordinary page via JavaScript, so the intruder can serve a page on localhost and open it in a normally launched Chrome instance without the debugging port or WebSocket client. The user-gesture can be satisfied by injecting an OS input event: XTest on Linux, the Win32 \texttt{SendInput} API on Windows, which the page receives as a fully trusted click. This grants user activation and starts the model download on both platforms. Input injection is itself a classic LotL technique.

Table~\ref{tab:gates} shows the outcomes of the user-activation checks applied before the model has been downloaded into the profile in use. A profile that already holds the model can create a headed or headless session immediately. A fresh profile correctly refuses without a gesture, but a single synthesized click is enough to satisfy the requirement. Furthermore, copying an existing install, the \texttt{OptGuideOnDeviceModel} directory together with the profile's \texttt{Local State} file, into a new profile yields a working model with no download or gesture. Copying the model directory alone does not bypass the gesture requirement.  

\begin{table}[t]
\centering
\footnotesize
\begin{tabular}{@{}lll@{}}
\toprule
Configuration & availability() & create() \\
\midrule
Existing profile, headed      & available    & ok \\
Existing profile, headless    & available    & ok \\
Fresh, no gesture     & downloadable & NotAllowedError \\
Fresh, CDP click      & downloadable & download starts \\
Fresh, OS-level click & downloadable & download starts \\
Fresh + model dir     & downloadable & NotAllowedError \\
Fresh + model + Local State & available  & ok \\
\bottomrule
\end{tabular}
\caption{Prompt API access under attacker-controlled launch configurations.
The OS-level click is a synthetic input event (XTest on Linux, Win32
\texttt{SendInput} on Windows), injected without the debugging protocol.
}
\label{tab:gates}
\end{table}

\subsection{From Model to Agent}

We drive the model with an agent harness (a simple tool-calling loop) written by the attacker as a PowerShell or Python script on the victim machine (Figure~\ref{fig:arch}). The harness keeps one session open for the whole task, so the model sees earlier tool results as conversation history. At each step, it prompts the model with a \texttt{responseConstraint} JSON schema that carries a free-text rationale, an action drawn from an enumeration of tool names, and the arguments that action needs. The constraint forces the model output to conform to the schema, but if a reply still fails to parse, the harness asks for a retry. This can happen when the context window is full.

The setup is deliberately flexible, and we tested a broad range of custom-defined tools written in PowerShell or Python: file operations (create, read, write, append, and list), arbitrary execution of commands, screenshot capture, image processing, raw keyboard and mouse manipulation, and GUI actuation through the accessibility tree. 

We found three design choices were necessary for such a small model. First, GUI actions are more accurate when using the accessibility tree rather than the model manually analyzing the screen from screenshots and manipulating the mouse. On Windows, \texttt{list\_ui\_elements} returns the named controls of the foreground window from UI Automation and \texttt{click\_element} clicks one by name at the OS reported coordinates. In early tests the model could not reliably locate controls from screenshots, often misjudging their positions while reporting confidence. Second, the session context window holds only 9,216 tokens, so we run image analysis through a separate session to get image descriptions and use an \texttt{append\_file} tool to keep intermediate results on disk rather than in context. Third, the system prompt steers the model toward command-line tools over GUI actions whenever both would work, due to the same limitations from above. The harness can also require operator confirmation of each action, allowing more control as needed.

\section{Evaluation}

\subsection{Setup}

We ran our experiments on a single laptop (RTX 2000 Ada GPU w/ 8GB VRAM; 32 threads) under Ubuntu 24.04 on WSL2 with Chrome 156 and on Windows 11 with Chrome 157 as a standard, non-elevated user with Chrome's process sandbox enabled. The model ran on the host GPU when available and fell back to the CPU otherwise, with identical outputs. On the headed Windows host, a prompt paged roughly 3GB of model weights into GPU memory and processed each prompt in roughly 1.4 s. Headless Chrome and the virtualized GPU under WSL2 both fall back to the CPU at around 3.7 s per prompt. The gating and policy outcomes we report were confirmed to be identical on both platforms. 

\subsection{Capability}\label{sec:capability}

We measure the model performance on tasks related to post-exploitation, including analyzing data on the host and carrying out multi-step operations.

\begin{figure*}[!t]
  \centering
  \includegraphics[width=\textwidth]{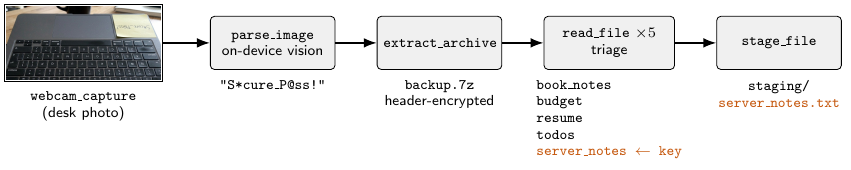}
  \caption{A simulated cyber operation, driven end-to-end by the on-device model. From a webcam photo of the desk the model reads the archive password off the sticky note ($\texttt{parse\_image}$), unlocks an encrypted archive ($\texttt{extract\_archive}$), triages the five extracted files, and copies out the one holding an SSH private key ($\texttt{stage\_file}$).}
  \label{fig:demo}
\end{figure*}

\begin{table}[b]
\centering
\footnotesize
\begin{tabular}{@{}lll@{}}
\toprule
Task & Corpus & Score \\
\midrule
Secret triage, hard neg.\ & CredData      & 0.77 \\
Secret triage, soft neg.\ & CredData$+$lib & 0.91 \\
Secret category           & CredData      & 0.68 \\
PII, any                  & Enron         & 0.92 \\
PII, sensitive subset     & Enron        & 0.74 \\
PII, per type             & Enron        & 0.68 \\
Image classification      & Imagenette   & 0.98 \\
\bottomrule
\end{tabular}
\caption{On-host data analysis performance.}
\label{tab:capability}
\end{table}

\paragraph{On-host data analysis.} Post-exploitation analysis includes finding credentials, triaging documents for value, and recognizing sensitive data whether in a file or an image. The browser-embedded model enables attackers to automate some of this analysis. We evaluate the model directly on public datasets, scoring each item against ground truth labels. We cover three tasks (Table~\ref{tab:capability}): credential triage in source code using CredData~\cite{creddata}, personal-data detection in real email on the Enron corpus~\cite{enron_klimt2004} with labels from a dedicated on-device PII detector~\cite{lfm2_pii_detector}, and image classification on Imagenette~\cite{imagenette} through the model's vision input capability.

We frame credential triage as a file-level detection task rather than verbatim extraction, as an attacker with host read access only needs to locate the credential-bearing files. In CredData, files without secrets are hard negatives that still contain secret-shaped decoys, such as configuration field names, default placeholder values, or variable references, making the task especially difficult for a small model. Using a balanced 500-file set, the model flags credential-bearing files with a recall of 0.93, rarely missing a real secret, but with a precision of only 0.65 ($F_1=0.77$), over-tagging realistic decoys like cryptographic API calls or redacted placeholders. A real host holds far more ordinary files than decoy-laden ones, so we also score against soft negatives: public library source files from the same CredData repos, but with no secrets or decoys. In this setting, precision rises to \textbf{0.90 at the same 0.93 recall ($\mathbf{F_1 = 0.91}$)}. We also evaluate the model's ability to categorize each secret as private keys, passwords, API tokens, auth credentials, or generic secrets. Assigning the credential categories to a flagged file reaches a micro-averaged $F_1$ of 0.68.

On the Enron corpus the model detects whether a message contains any personal data with
$\mathbf{F_1 = 0.92}$, and the sensitive subset of contact details, financial data, and government identifiers with
$F_1 = 0.74$. Classifying PII into types reaches a micro-averaged $F_1$ of 0.68, strongest on names,
email addresses, and phone numbers ($0.77$--$0.92$), with the sparsely labeled financial
and government-identifier types reported as lower-confidence. 

For vision, on the Imagenette validation set (10 classes, 250 images balanced at 25 per class) the model
reaches \textbf{98.0\% top-1 accuracy} with per-class recall never below 0.92, with errors being semantically plausible confusions. 

\paragraph{Tool-call accuracy.} The model must be capable of emitting well-formed tool calls to be used as an agent for carrying out multi-step operations. We evaluate tool-calling ability with the Berkeley Function-Calling Leaderboard (BFCL)~\cite{bfcl}, using its single-turn categories and official abstract-syntax-tree checker (Table~\ref{tab:toolcall}). On each item, the model receives the candidate function schemas and the user query and is constrained to emit exactly one JSON call, or to abstain. When given a single function schema, it correctly filled in the arguments and called the function 90\% of the time. When given multiple functions, the model selected and correctly called the function with an \textbf{accuracy of 92\%}. The small model struggles the most in cases where no function applies to the query, correctly abstaining 75\% of the time and otherwise inventing a plausible-looking call. Aggregating across all three settings resulted in an accuracy of 86\%.

\begin{table}[b]
\centering
\footnotesize
\begin{tabular}{@{}llll@{}}
\toprule
Category & $n$ & What it tests & Accuracy \\
\midrule
Simple      & 400 & one function, fill arguments     & 0.90 \\
Multiple    & 200 & pick the right function          & 0.92 \\
Irrelevance & 240 & abstain; no function applies     & 0.75 \\
\midrule
Overall     & 840 &                                  & 0.86 \\
\bottomrule
\end{tabular}
\caption{Tool-calling accuracy.}
\label{tab:toolcall}
\end{table}

Overall, the browser-embedded model performance is a far-cry from large frontier models, but its ability to triage files, classify images, and use tools is good enough to aid an attacker in post-exploitation activities. We demonstrate this next by constructing a simulated cyber operation that requires the multi-step use of these skills.

\paragraph{Multi-step execution.} To demonstrate these primitives composed into an operation, we built a single multi-step case study, driven end-to-end by the on-host model. The scenario chains the capabilities measured above and crosses from the physical disk to the credential store: a password written on a sticky note decrypts an archive containing multiple files on the host, one of which leaks an SSH private key. We wrote a simple PowerShell agent harness (and a Python equivalent) exposing six tools: \texttt{webcam\_capture} (injects a synthetically generated image), \texttt{parse\_image}, \texttt{list\_dir}, \texttt{read\_file}, \texttt{extract\_archive}, and \texttt{stage\_file}, with the model reached over a localhost page as our access technique. The target is a header-encrypted \texttt{7z} archive whose filenames are hidden until the password is recovered. Inside are five files, one holding the private key and four benign (book notes, a budget, a resume, and a todo file), so locating
the credential is a realistic triage task (Figure~\ref{fig:demo}).

The goal trajectory is to use a \texttt{webcam\_capture} to get the sticky note from the desk, \texttt{parse\_image} to read the password off the note, \texttt{extract\_archive} to unlock the 7-zip, \texttt{read\_file} over the contents of each file in the archive, and \texttt{stage\_file} on the one file bearing the key. We found the decisive factor is how the goal is presented. Given an open-ended objective (``recover any credentials''), the model completed the operation in only 60\% of trials: it reliably executed each step but planned poorly across them, failing to recognize when a found credential should be staged and burning its step budget ($n=20$) on redundant actions. Given the same goal as an explicit numbered plan (Figure~\ref{fig:plan}), with the harness maintaining a step checklist (each step marked done or pending) as a working memory, the model successfully completed the operation in $41$ of $50$ trials. In $2$ of the other $9$ trials, the model completed the entire chain up to identifying the SSH key, but staged the todo.txt file instead of the server notes. The other $7$ trials exhausted the step budget. On inspection, the model struggled with parsing the password correctly from the image, and would repeat the step after failing to decrypt the archive. All $7$ eventually recovered the correct password, but without enough budget to complete the rest of the operation. Combined, the results suggest the browser-embedded model is an unreliable autonomous planner, but a reliable executor of an operator supplied plan, a realistic LotL use case for an attacker. The operation also inherits the technique's stealth, since the model runs on a loopback page with no inference-time network egress, the archive is opened and the key staged entirely
on the host, and the model calls and compute all done on-device.

\begin{figure}[t]
\centering
\begin{lstlisting}[basicstyle=\scriptsize\ttfamily,frame=single,numbers=none,
  columns=fullflexible,breaklines=true,xleftmargin=4pt,xrightmargin=2pt]
GOAL: recover the SSH private key kept in the
encrypted backup, and copy the file that contains
it into the staging folder.

PLAN:
  1. Photograph the desk with the webcam and read the
     archive password off the sticky note.
  2. Extract the archive.
  3. Read the extracted files and find the one that
     contains a credential.
  4. Copy that one credential-bearing file into the
     staging folder.
\end{lstlisting}
\caption{The numbered plan given to the model. Each turn the harness also supplies a checklist that marks every step done or pending as working memory.}
\label{fig:plan}
\end{figure}

\subsection{Stealth and Visibility}

As mentioned, our trials produced no network traffic to observe, therefore a security product is limited to the agent's on-host behavior and analysis of the written harness. Credential, PII, and other document triage requires only file reads, and the other capabilities are similarly achieved using existing host utilities. A more feature-rich version of the harness was flagged by Windows Defender, as the ability to capture screenshots and move the mouse made the harness resemble a known post-exploitation kit, \emph{PowerShell Empire}. Our smaller, operation focused harness evaded detection, and would be easy for an attacker to write on-the-fly instead of requiring a download or transfer over the network.  


\section{Risks and Implications}
A browser-embedded model changes the economics of post-exploitation in three ways. First, it lets an intruder search and summarize victim data on the host at a much greater speed and only a small result ever needs to leave the network. Second, it removes the revocable choke points of remote LM use: there is no API key to cancel, no provider-side abuse monitoring, and no traffic to block if the model is already downloaded. Third, its reach scales with the browser install base on hardware that meets the model's requirements, which already includes typical laptops and PCs. The pattern is not specific to Chrome; any platform that ships a local model to general-purpose endpoints creates a similar asset for attackers.

The main limit today is the small model's error rate, which we expect will improve over time. Chrome is currently on version 3 of Nano, and if trends continue, the on-device model will improve in future iterations. Vendors are directly incentivized to make the model better at reading, summarizing, and acting on web content, and these are the same capabilities an attacker repurposes to triage on-system data. Improvements shipped for the benign use case accrue to the malicious one at no cost to the attacker. 

\section{Mitigations and Detection}

\begin{table}[t]
\centering
\footnotesize
\begin{tabular}{@{}>{\raggedright\arraybackslash}p{0.3\columnwidth}>{\centering\arraybackslash}p{0.11\columnwidth}>{\centering\arraybackslash}p{0.11\columnwidth}>{\raggedright\arraybackslash}p{0.32\columnwidth}@{}}
\toprule
Policy (setting) & Stops CDP & Stops page & Cost to legit use \\
\midrule
BuiltInAI\-APIs\-Enabled (false) & yes & yes & Breaks built-in AI web apps \\
GenAI\-Local\-Foundational\-Model\-Settings (1) & yes & yes & Deletes model; removes on-device features \\
RemoteDebugging\-Allowed (false) & yes & \textbf{no} & Breaks CDP automation \\
DeveloperTools\-Availability (2) & yes & \textbf{no} & Breaks DevTools \\
\bottomrule
\end{tabular}
\caption{Enterprise policies against two ways of driving the on-device model.}
\label{tab:policies}
\end{table}

Because Chrome's threat model excludes code that already runs as the user, the browser does not treat on-device model access as a security boundary~\cite{chromium_security_faq}. Protection therefore rests with the organization: what it allows through enterprise policy, and what it can observe on the endpoint.

\paragraph{Policy.} Table~\ref{tab:policies} summarizes four Chrome enterprise policies impacting the model. Only the two that govern the model itself fully prevent its use. Setting \texttt{BuiltInAI\allowbreak APIs\allowbreak Enabled} to \texttt{false} makes every built-in API report the model as unavailable, including in freshly created profiles, and setting
\texttt{GenAILocal\allowbreak Foundational\allowbreak Model\allowbreak Settings} to \texttt{1} prevents the download and deletes an existing model~\cite{chromium_policy_builtin_ai,chromium_policy_genai_local}. The two debugging policies, \texttt{Remote\allowbreak Debugging\allowbreak Allowed} and \texttt{Developer\allowbreak Tools\allowbreak Availability}, block only our DevTools-based harness and not model use in general, so neither should be relied on as a control~\cite{chromium_policy_remote_debugging}. None of the four is enabled by default in a fresh install. These policies carry a real cost, since they also disable legitimate built-in features and any internal web applications built on them. We recommend that individuals and organizations make an explicit decision rather than inherit the default. Users with no need for on-device models should disable them, and others should treat the model as a dual-use capability, the same way they treat PowerShell or WMI, and make efforts toward detecting malicious use.

\paragraph{Detection.} On-device inference produces no network traffic, so detection has
to rely on host telemetry. The signals we observed fall into three groups:
\begin{itemize}
  \item \emph{Process lineage.} Chrome launched by a script host such as Python or
    PowerShell rather than by the shell or a shortcut, with a non-default
    \texttt{--user-data-dir}, and often in headless mode. A remote-debugging switch on
    that command line is a strong additional indicator.
  \item \emph{Local listeners.} The parent script holding a loopback listener that the
    Chrome child connects to shortly after launch.
  \item \emph{Model files out of place.} \texttt{OptGuideOnDeviceModel} or
    \texttt{Local State} written into a profile directory that never downloaded the
    model, i.e.\ an install copied from another profile.
\end{itemize}

The only host-visible behaviors are the agent's actuators, like input injection and screen
capture common to other post-exploitation tooling, so existing behavioral detections may catch
those but not the model's reasoning. Analysis-only use offers even less to see, as a combination of file reads and model calls. The signals above therefore target the attacker intentions directly. The process-lineage signal in particular, Chrome launched by a script host, pointed at its own profile, with a remote-debugging or headless switch, can be expressed as a process-creation detection, though we offer these as proposed signals rather than ones validated against production logs.

\paragraph{Platform recommendations.} Browser vendors could narrow the gap without changing their threat model: expose model-session events through enterprise reporting so defenders can audit which origins use the model and how often; surface model use to the user, as browsers already do for the camera and microphone; and bind a downloaded model to the profile that downloaded it, so an install cannot simply be copied into a new profile.

\section{Conclusion}

Browser-embedded language models hand intruders a capable, offline post-exploitation engine that is already installed and signed by a trusted vendor. We showed that Chrome's Gemini Nano can be reached from the host using only shipped interpreters, that its one-time user-activation gate is satisfied by a synthesized click or sidestepped by copying an existing install, and that the model is competent enough to triage credential-bearing files, flag personal data, classify images, call tools, and compose these primitives into a complete operation, all with no inference-time traffic for a defender to observe. The capability is neither unique to Chrome nor likely to regress: any platform that ships a local model to general-purpose endpoints creates the same asset, and the vendor improvements that make these models better at reading and acting on content accrue directly to the attacker.

Organizations should therefore inventory and govern on-device models the way they govern any other dual-use capability such as PowerShell or WMI: decide deliberately whether to enable them, disable them where they serve no purpose, and otherwise watch the host-level signals we describe for signs of abuse. Browser vendors, in turn, can narrow the gap without changing their threat model, by surfacing model use to the user as they already do for the camera and microphone and binding a downloaded model to each profile.

\bibliography{references}

\end{document}